\documentclass[emulateapj,twocolumn,times,compact]{aastex62}

\definecolor{red}{rgb}{1.0,0.0,0.0}
\graphicspath{{./}{figures/}}
\received{2018 December 11}
\revised{2019 January 17}
\accepted{2019 January 21}
\submitjournal{AJ}
\shorttitle{A Near-coplanar Stellar Flyby of the Planet Host Star HD 106906}
\shortauthors{De Rosa \& Kalas}

\begin{document}

\title{A Near-coplanar Stellar Flyby of the Planet Host Star HD 106906}

\correspondingauthor{Robert J. De Rosa}
\email{rderosa@stanford.edu}

\author[0000-0002-4918-0247]{Robert J. De Rosa}
\affiliation{Kavli Institute for Particle Astrophysics and Cosmology, Stanford University, Stanford, CA 94305, USA}
\affil{Department of Astronomy, University of California, Berkeley, CA 94720, USA}

\author[0000-0002-6221-5360]{Paul Kalas}
\affil{Department of Astronomy, University of California, Berkeley, CA 94720, USA}
\affil{SETI Institute, 189 Bernardo Ave., Mountain View, 94043, USA}

\begin{abstract}
We present an investigation into the kinematics of HD 106906 using the newly released {\it Gaia} DR2 catalog to search for close encounters with other members of the Scorpius--Centaurus (Sco--Cen) association. HD 106906 is an eccentric spectroscopic binary that hosts both a large asymmetric debris disk extending out to at least 500\,au and a directly imaged planetary-mass companion at a projected separation of 738\,au. The cause of the asymmetry in the debris disk and the unusually wide separation of the planet is not currently known. Using a combination of {\it Gaia} DR2 astrometry and ground-based radial velocities, we explore the hypothesis that a close encounter with another cluster member within the last 15\,Myr is responsible for the present configuration of the system. Out of 461 stars analyzed, we identified two candidate perturbers that had a median closest approach (CA) distance within 1\,pc of HD 106906: HIP 59716 at $D_{\rm CA}=0.65_{-0.40}^{+0.93}$\,pc ($t_{\rm CA}=-3.49_{-1.76}^{+0.90}$\,Myr) and HIP 59721 at $D_{\rm CA}=0.71_{-0.11}^{+0.18}$\,pc ($t_{\rm CA}=-2.18_{-1.04}^{+0.54}$\,Myr), with the two stars likely forming a wide physical binary. The trajectories of both stars relative to HD 106906 are almost coplanar with the inner disk ($\Delta\theta = 5\fdg4\pm1.7$ and $4\fdg2_{-1.1}^{+0.9}$). These two stars are the best candidates of the currently known members of Sco--Cen for having a dynamically important close encounter with HD 106906, which may have stabilized the orbit of HD 106906 b in the scenario where the planet formed in the inner system and attained high eccentricity by interaction with the central binary.
\end{abstract}


\section{Introduction}
\label{sec:intro}
Close encounters by external stellar perturbers have the potential to significantly modify the architecture of planetary systems \citep{laughlin98a, kenyon04a}. Stars passing very close to our solar system have been invoked to explain the formation of the Oort cloud of comets \citep{duncan87a}, comet showers \citep{hills81a}, the disruption of the Kuiper Belt \citep{kobayashi05a}, and the distant detached orbits of dwarf planets such as 90377 Sedna \citep{brown04a}, as well as the hypothetical Planet Nine \citep{Bromley:2016kg}.  Such close stellar encounters have also been invoked to explain the orbital properties of extrasolar planets \citep{zakamska04a, malmberg09a, Parker:2011bd, Pfalzner:2018hi}, as well as the observed structure of disks around \object{HD 100453} \citep{Wagner:2018dg}, \object{HD 141569} \citep{Ardila:2005gd,Reche:2008fm},  \object{HD 15115} \citep{kalas07a}, and \object[bet Pic]{$\beta$~Pictoris} \citep{Kalas:1995jo,Ballering:2016ih}. 

In the case of $\beta$~Pic, \citet{kalas00a} and \citet{Larwood:2001ks} explore numerical models of a non-coplanar stellar flyby ($\Delta i=30^\circ$) that qualitatively reproduce the disk asymmetries measured beyond the $\sim$200\,au projected radius; the northeast disk extension is radially extended and vertically flat whereas the southwest disk extension is radially truncated and vertically extended \citep{Kalas:1995jo}.  Using astrometry from the {\it Hipparcos} catalog, \citet{Kalas:2001dp} identified several candidate perturbers that had a close encounter with the $\beta$~Pic system in the last 1\,Myr that could explain the asymmetry. The proximity and large apparent motion of $\beta$~Pic were conducive for this analysis given the precision of the measurements available at the time. The order of magnitude improvement in the astrometric precision of the recently published {\it Gaia} DR2 catalog \citep{GaiaCollaboration:2018io} makes it feasible to extend this analysis to more distant stars that also show evidence of dynamical perturbation.

In this paper we present an investigation into the dynamical history of \object{HD 106906}, an F5V \citep{Houk:1975vw} member of the 15\,Myr-old \citep{Pecaut:2016fu} Lower Centaurus--Crux subgroup of the Scorpius--Centaurus OB association \citep{Chen:2011gz}. HD 106906 rose to prominence with the discovery of a planetary-mass companion ($11\pm2$ M$_{Jup}$) at a wide projected separation of 738\,au \citep{Bailey:2014et}, exterior to a debris disk that was inferred from the spectral energy distribution of the star \citep{Chen:2005gs}. The inner disk was later resolved in near-infrared scattered light as a $\sim$50\,au radius ring inclined to the line of sight by $\sim$85$^\circ$ \citep{Kalas:2015en,Lagrange:2016bh}.  However, visible light {\it Hubble Space Telescope} ({\it HST}) data showed a striking asymmetry in the morphology of the outer disk that resembles the disk asymmetry of $\beta$ Pic \citep{Kalas:2015en}. The planet is oriented $\sim$21$^{\circ}$ from the position angle of the disk midplane, suggesting that the planet's orbit is not coplanar with the disk \citep{Kalas:2015en}. HD 106906 is also a spectroscopic binary with a mass ratio near unity \citep{Lagrange:2019}. 

The hypothetical dynamical history of this system has been further studied by \citet{Jilkova:2015jo}, \citet{Nesvold:2017ho}, and \citet{ Rodet:2017hr}. \citet{ Rodet:2017hr} quantified a scenario where \object{HD 106906 b} originally formed in a disk near the binary star, migrated inward, encountered an unstable mean-motion resonance with the binary, was ejected into a high-eccentricity orbit, and then had its periastron raised into a stable region by an external stellar perturber. Here we search for potential stellar perturbers consistent with this scenario using the exquisite precision provided by the {\it Gaia} DR2 astrometry and ground-based radial velocities measurements.

\section{Systemic Velocity of HD 106906}
HD 106906 consists of a $1.37\,M_{\odot}$ primary and $1.34\,M_{\odot}$ secondary with a projected separation of between 0.36--0.58\,au \citep{Rodet:2017hr}. The full orbital characterization of the system---most critically the systemic velocity---has not yet been published. There are four radial velocities for this system within the literature, two are instantaneous ($10.2\pm1.7$\,km\,s$^{-1}$, \citealp{Gontcharov:2006dn}; $15.1\pm0.3$\,km\,s$^{-1}$, \citealp{Chen:2011gz}), and two are from a combination of multiple measurements ($8.4$\,km\,s$^{-1}$, $N=22$, \citealp{Evans:1967vu}; $11.1$\,km\,s$^{-1}$, $N=2$, \citealp{Nordstrom:2004ci}). Given the scatter in these measurements, possibly caused by the orbit of the binary, we opted instead to use archival spectroscopic observations of HD 106906 to measure the systemic velocity.

\begin{figure}
\epsscale{1.15}
\plotone{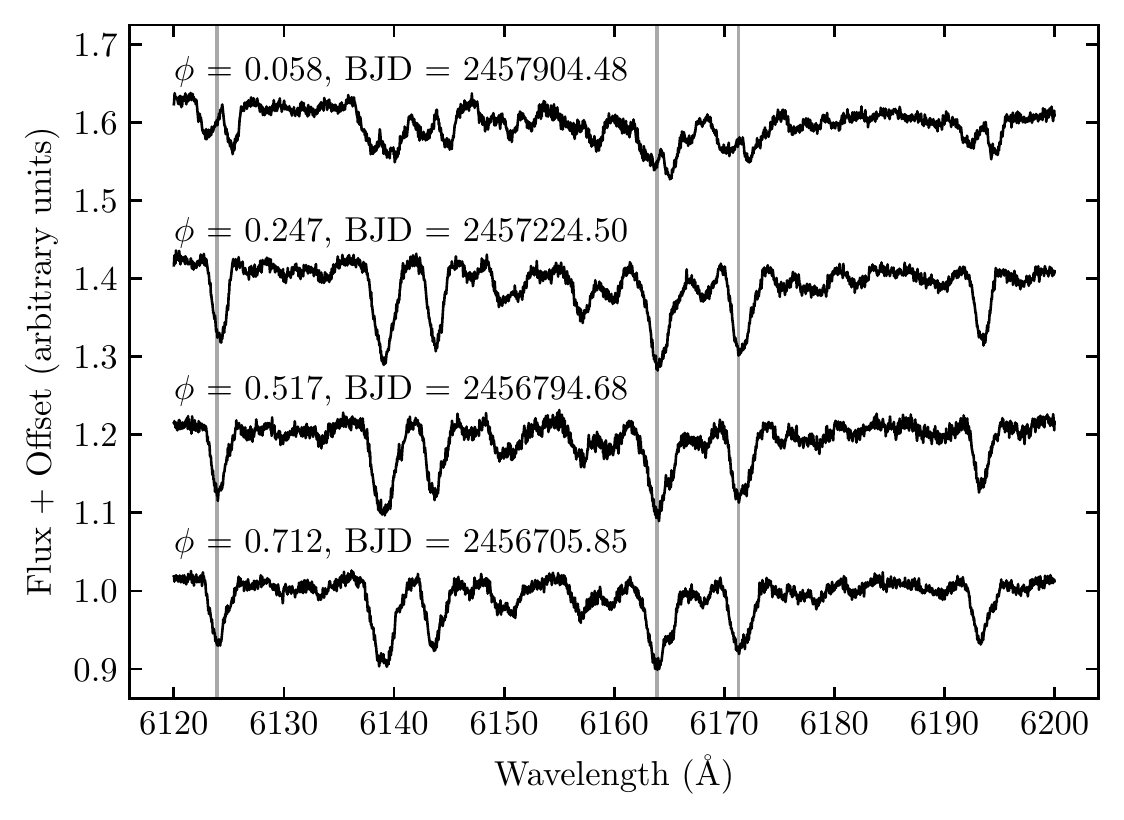}
\caption{HARPS spectra of the HD 106906 binary system showing significant changes in line morphology due to the relative velocities of the two components. The four spectra cover a range of different orbital phases: near periapsis ($\phi=0.058$) where the velocity differential is highest and the lines of the two components are resolved, nine days later in the orbit ($\phi=0.247$) where the spectra lines are the narrowest, and two subsequent phases where the shallower lines of the primary are being red-shifted, distorting the line profile. The rest wavelengths of three calcium lines are indicated.}
\label{fig:1}
\end{figure}

Calibrated optical spectra of HD 106906 from the High Accuracy Radial velocity Planet Searcher (HARPS; \citealp{Mayor:2003wv}) instrument were obtained from the European Southern Observatory (ESO) Archive (program IDs 192.C-0224, 098.C-0739, 099.C-0205) for 76 epochs from 2014 January  to 2017 June. The radial velocities computed automatically by the HARPS Data Reduction Software were strongly biased by the blending of the rotationally broadened lines of both stars in all but four of the epochs, and were therefore not used in the subsequent analysis. Instead, we use the epoch in which the lines of the two stars are the most separated (2017 May 30) to fit a model atmosphere \citep{Husser:2013ca} to both stars which can then be used as a template to fit the Keplerian motion of both stars in the remaining epochs (Figure \ref{fig:1}). The fit was restricted to seven lines/groups of lines centered on 5040, 5370, 5448, 5529, 6141, 6441, and 6680 \AA. The best-fit model was found via $\chi^2$ minimization by varying the temperature, surface gravity, radial velocity and rotational velocity of both stars, the flux ratio, and the metallicity. For each trial the model atmosphere was generated via linear interpolation, rotationally broadened, Doppler shifted, and smoothed to the resolution of HARPS. The HARPS and model atmosphere spectra were then continuum normalized with a linear fit to the continuum.

The resulting template for each star was used to fit the Doppler shift induced by the Keplerian motion of each star over the three-year baseline. We used the affine-invariant Markov chain Monte Carlo package \texttt{emcee} \citep{ForemanMackey:2013io} to fit the seven orbital elements: period $P$, eccentricity $e$, argument of periapsis $\omega$, epoch of periapsis $\tau$, primary semi-amplitude $K_1$, mass ratio $q$, and the systemic velocity $\gamma$. At each step in the chain, the radial velocity of both stars at each epoch was predicted from the orbital elements. The templates constructed previously were Doppler shifted to the corresponding velocity, summed, and compared to the observed spectrum at that epoch. Mass functions that were significantly discrepant with the apparent brightness of the system were excluded. The period ($P=49.233\pm0.001$\,days) and eccentricity ($e=0.669\pm0.002$) were well constrained, and are consistent with an independent analysis by \citet{Lagrange:2019}. We found a systemic velocity of $\gamma=12.18\pm0.15$\,km\,s$^{-1}$ after marginalizing over the remaining parameters, the error being a combination of statistical error from the fit and an assumed systematic uncertainty to account for biases in the fitting process estimated by repeating the fitting process using subsets of the lines described previously.

\section{Close stellar encounters with HD 106906}
\label{sec:linear}
\subsection{Identifying candidate perturbers}
The revised systemic radial velocity was combined with astrometry from {\it Gaia} DR2 \citep{GaiaCollaboration:2018io} to compute the position and kinematics of HD 106906 given in Table~\ref{tab:1}. Uncertainties were propagated in a Monte Carlo fashion, drawn from a multivariate normal distribution using the correlation coefficients given in the {\it Gaia} catalog for the astrometry, and from a normal distribution for the radial velocity. One potential source of error in the {\it Gaia} astrometry is the motion of the photocenter relative to the center of mass. All sources within the {\it Gaia} catalog were fit with a five-parameter single star model. The magnitude of this error should be small given the comparable mass and flux ratio of the two components.

A catalog of Sco--Cen members was constructed following \citet{Wright:2018da}, augmented with members reported in \citet{Song:2012gc}. Astrometric measurements were obtained from {\it Gaia} DR2, while radial velocities came from a number of sources, either from various literature sources \citep{Levato:1996ta, Torres:2006bw, Gontcharov:2006dn, Kharchenko:2007eh, Chen:2011gz, Song:2012gc, Dahm:2012iu, Kordopatis:2013cd, Desidera:2015fl}, or from the {\it Gaia} catalog. The final sample consisted of 461 members that had both a {\it Gaia} DR2 parallax and either a ground-based or {\it Gaia} radial velocity measurements. 

Candidate perturbers were identified through a linear approximation of the space motion of each of the sample stars relative to HD 106906. In this approximation, the velocity of each star is constant and gravitational forces are not considered. Uncertainties in the true position and motion of each star were propagated in a Monte Carlo fashion. A trial position, proper motion, and parallax were drawn from a multivariate normal distribution using the correlation coefficients in the {\it Gaia} catalog, and a radial velocity was drawn from a normal distribution. Closest approach times and distances ($D_{\rm CA}$) were computed for each of the $10^5$ trials. This process was repeated for each star within the sample.

\subsection{HIP 59716 and HIP 59721}
\startlongtable
\begin{deluxetable*}{rcccc}
\tabletypesize{\footnotesize}
\tablecaption{Kinematics for HD 106906 and two candidate perturbers}
\tablecolumns{5}
\tablenum{1}
\tablewidth{0pt}
\tablehead{\colhead{Property} & \colhead{Unit} & \colhead{HD 106906} & \colhead{HIP 59716} & \colhead{HIP 59721}}
\startdata
$\alpha$ & deg & $184.47133075613\pm (1.48\times10^{-8})$ & $183.71107188480\pm(1.31\times 10^{-8})$ & $183.71768277794\pm(1.27\times 10^{-8})$ \\
$\delta$ & deg & $-55.97558054009\pm (7.41\times10^{-9})$ & $-55.78991603504\pm(6.41\times 10^{-9})$ & $-55.78439084834\pm(5.86\times 10^{-9})$ \\
$\pi$ & mas & $9.6774\pm0.0429$ & $8.7068\pm0.0393$ & $8.6815\pm0.0349$ \\
$\mu_{\alpha} \cos\delta$ & mas\,yr$^{-1}$ & $-39.014\pm0.057$ & $-35.722\pm0.050$ & $-35.184\pm0.048$ \\
$\mu_{\delta}$ & mas\,yr$^{-1}$ & $-12.872\pm0.046$ & $-11.041\pm0.042$ & $-11.325\pm0.039$\\
RV & km\,s$^{-1}$ & $12.18\pm0.15$\tablenotemark{a} & $15.5\pm1.1$\tablenotemark{b} & $17.6\pm1.7$\tablenotemark{c}\\
\hline
\multicolumn{5}{c}{Galactic position and motion}\\
\hline
$X$ & pc & $48.54\pm0.22$ & $53.14\pm0.24$ & $53.30\pm0.22$ \\
$Y$ & pc & $-90.45\pm0.40$ & $-100.93\pm0.46$ & $-101.22\pm0.41$ \\
$Z$ & pc & $11.86\pm0.05$ & $13.43\pm0.06$ & $13.48\pm0.05$ \\
$U$ & km\,s$^{-1}$ & $-9.77\pm0.10$ & $-8.63\pm0.51$ & $-7.42\pm0.79$ \\
$V$ & km\,s$^{-1}$ & $-20.12\pm0.14$ & $-23.09\pm0.97$ & $-24.83\pm1.49$ \\
$W$ & km\,s$^{-1}$ & $-7.29\pm0.05$ & $-6.85\pm0.14$ & $-6.74\pm0.20$ \\
\hline
\multicolumn{5}{c}{Galactocentric position and motion}\\
\hline
$X_G$ & pc & $-8251.38\pm0.22$ & $-8246.77\pm0.24$ & $-8246.61\pm0.22$ \\
$Y_G$ & pc & $-90.45\pm0.40$ & $-100.93\pm0.46$ & $-101.22\pm0.41$ \\
$Z_G$ & pc & $38.70\pm0.05$ & $40.26\pm0.06$ & $40.31\pm0.05$ \\
$U_G$ & km\,s$^{-1}$ & $1.30\pm0.10$ & $2.45\pm0.51$ & $3.66\pm0.79$ \\
$V_G$ & km\,s$^{-1}$ & $212.12\pm0.14$ & $209.15\pm0.97$ & $207.41\pm1.49$ \\
$W_G$ & km\,s$^{-1}$ & $-0.01\pm0.05$ & $0.43\pm0.14$ & $0.54\pm0.20$ \\
\hline
\multicolumn{5}{c}{HD 106906 rest frame ($-X^{\prime}$ toward Earth, $Y^{\prime}$ toward East, $Z^{\prime}$ toward North)}\\
\hline
$X^{\prime}$ & pc & \nodata & $11.52\pm0.69$ & $11.85\pm0.65$ \\
$Y^{\prime}$ & pc & \nodata & $-0.8568\pm0.0039$ & $-0.8519\pm0.0034$ \\
$Z^{\prime}$ & pc & \nodata & $0.3675\pm0.0017$ & $0.3797\pm0.0015$ \\
$U^{\prime}$ & km\,s$^{-1}$ & \nodata & $3.19\pm1.11$ & $5.30\pm1.71$ \\
$V^{\prime}$ & km\,s$^{-1}$ & \nodata & $-0.39\pm0.13$ & $-0.16\pm0.13$ \\
$W^{\prime}$ & km\,s$^{-1}$ & \nodata & $0.13\pm0.06$ & $-0.03\pm0.06$ \\
\hline
\multicolumn{5}{c}{HD 106906 debris disk frame (disk lies in $Y_{\rm disk}Z_{\rm disk}$ plane)}\\
\hline
$X_{\rm disk}$ & pc           & \nodata & $1.136\pm0.061$ & $1.179\pm0.057$ \\
$Y_{\rm disk}$ & pc           & \nodata & $-0.9227\pm0.0042$ & $-0.9212\pm0.0038$ \\
$Z_{\rm disk}$ & pc           & \nodata & $-11.46\pm0.69$ & $-11.79\pm0.64$ \\
$U_{\rm disk}$ & km\,s$^{-1}$ & \nodata & $0.30\pm0.13$ & $0.39\pm0.17$ \\
$V_{\rm disk}$ & km\,s$^{-1}$ & \nodata & $-0.41\pm0.12$ & $-0.15\pm0.11$ \\
$W_{\rm disk}$ & km\,s$^{-1}$ & \nodata & $-3.18\pm1.10$ & $-5.29\pm1.70$ \\
\hline
\enddata
\tablenotetext{a}{This work}
\tablenotetext{b}{\citet{Chen:2011gz}}
\tablenotetext{c}{\citet{Song:2012gc}}
\label{tab:1}
\end{deluxetable*}
\begin{figure*}
\plotone{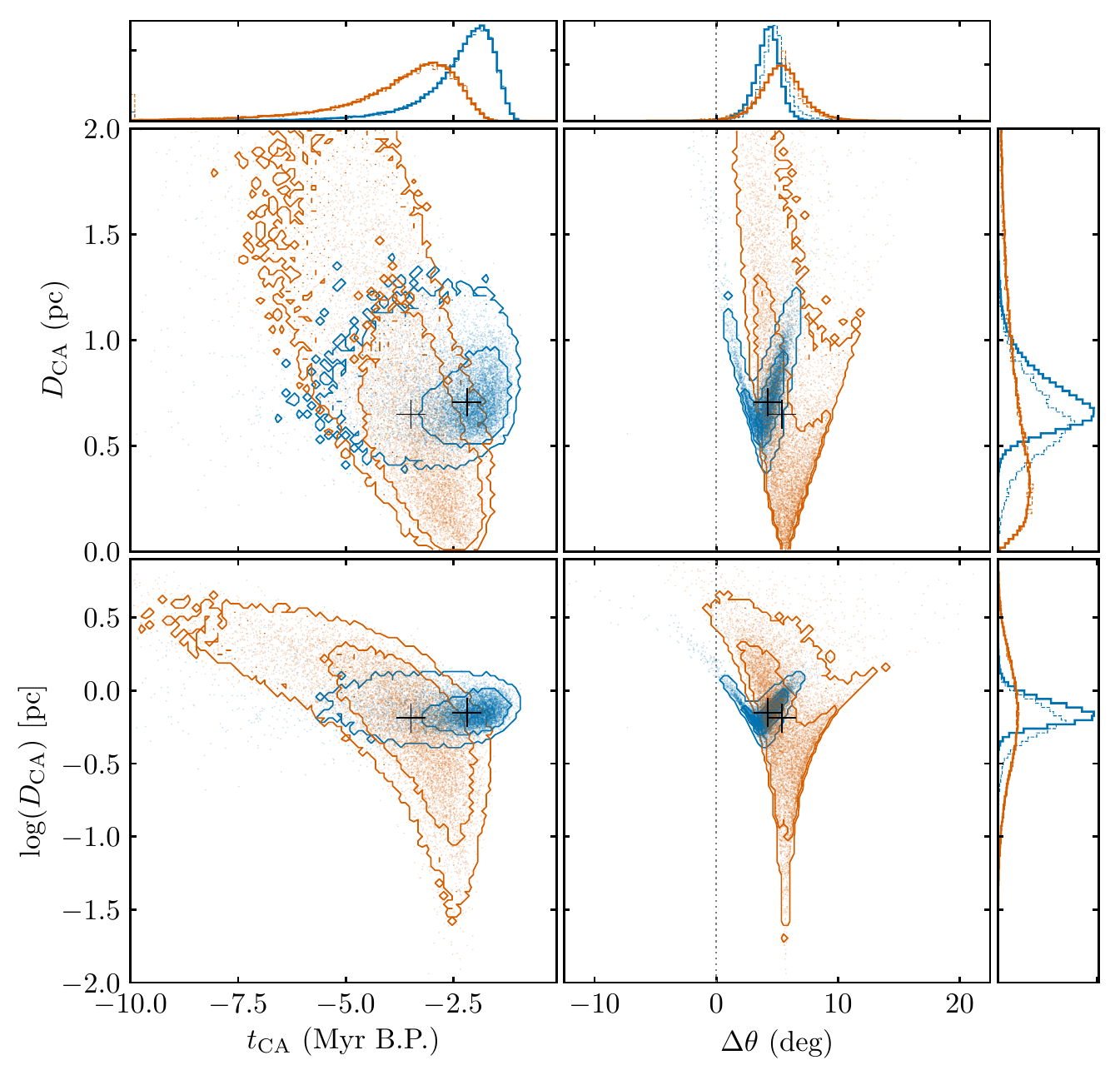}
\caption{Closest approach distance (left) and the angle between the trajectory of the flyby and the plane of the disk (right) as a function of time before present for the two candidate perturbers that had a median closest approach distance within 1\,pc, HIP 59716 (red) and HIP 59721 (blue). The contours and solid histograms are from the linear approximation. Contours denote the $1\sigma$ and $2\sigma$ credible regions. For clarity, only the marginalized distributions are shown for the results of the $N$-body simulation (light histogram).}
\label{fig:2}
\end{figure*}
\begin{figure}
\epsscale{1.15}
\plotone{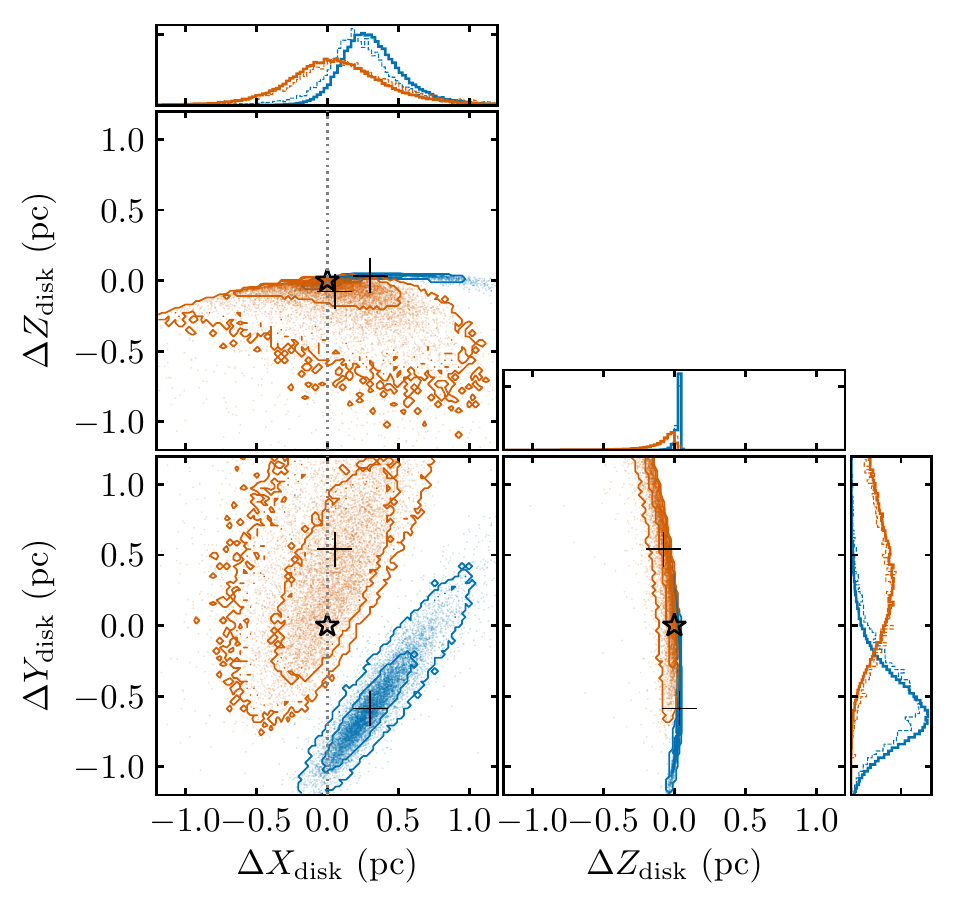}
\caption{Closest approach locations of HIP 59716 (red) and HIP 59721 (blue) relative to HD 106906 (black star) from the linear approximation in the coordinate system defined by the HD 106906 disk. The correlation between pairs of coordinates are shown, as well as marginalized distributions. The cross indicates the location of the 50th percentile in the marginalized distributions; the size of the symbol is arbitrary. Solid contours denote 1 and 2-$\sigma$ credible regions. The corresponding marginalized distributions from the $N$-body simulation of the three stars are also plotted (light histograms).}
\label{fig:xyz}
\end{figure}
\begin{figure*}
\plottwo{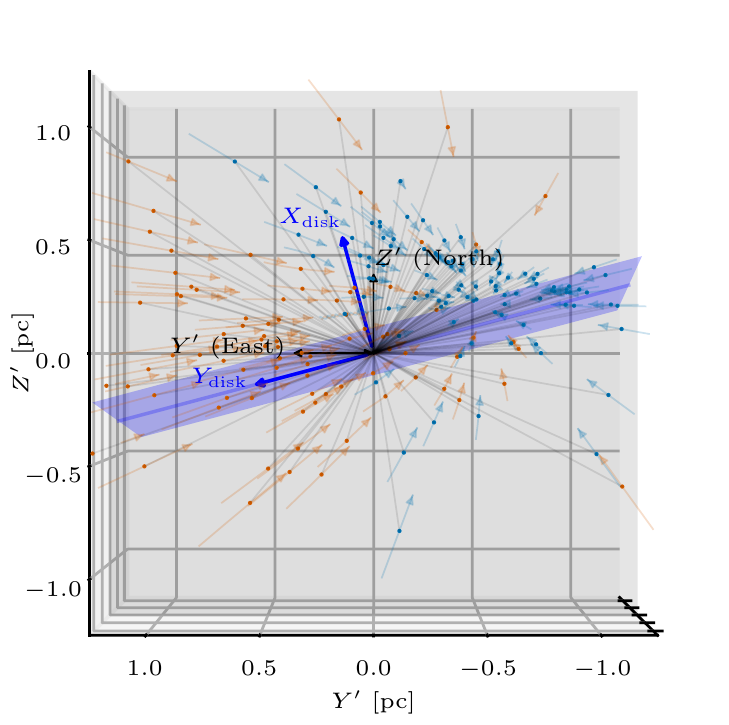}{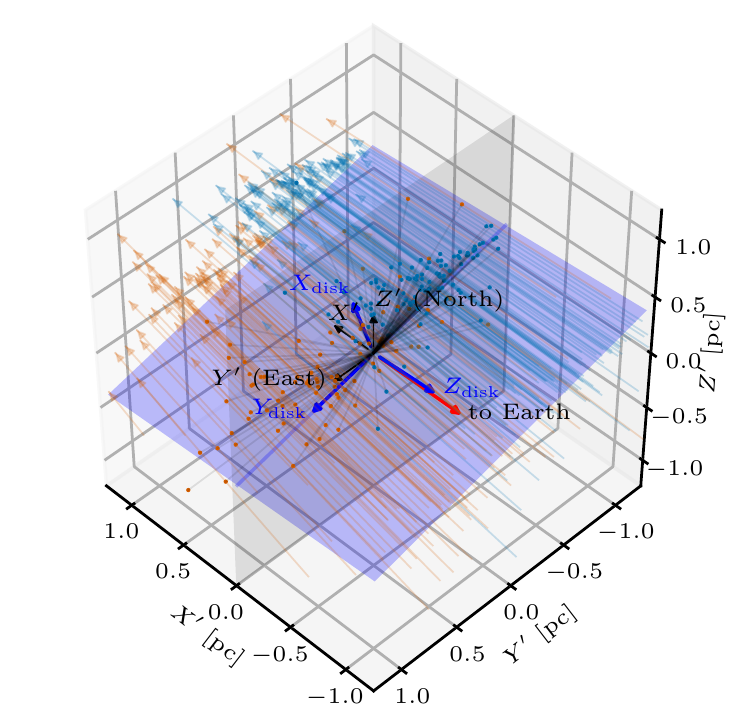}
\caption{Schematic of the close encounter between HD 106906, HIP 59716 (red tracks), and HIP 59721 (blue tracks) in the rest frame of HD 106906. The left panel shows the flyby as viewed from the Sun's current position relative to HD 106906, and the right panel is rotated $45^{\circ}$ in both azimuth and elevation. In both panels the sky plane ($Y^{\prime}Z^{\prime}$) is denoted by a gray plane, and the disk plane ($Y_{\rm disk}Z_{\rm disk}$) by a blue plane. The location of closest approach for each trajectory is marked by a point connected to the origin with a gray line. Both stars approach HD 106906 from the $-X^{\prime}$ direction, moving almost entirely in the $X^{\prime}Y^{\prime}$ plane.}
\label{fig:3}
\end{figure*}

Two candidate perturbers, likely forming a wide physical binary, were identified as having a closest approach distance within 1\,pc of HD 106906 within the last 15\,Myr: \object{HIP 59716} (F5V, \citealp{Houk:1975vw}) at $t_{\rm CA}=-3.49_{-1.76}^{+0.90}$\,Myr, $D_{\rm CA}=0.65_{-0.40}^{+0.93}$\,pc; and \object{HIP 59721} (G9V, \citealp{Torres:2006bw}) at $t_{\rm CA}=-2.18_{-1.04}^{+0.54}$\,Myr, $D_{\rm CA}=0.71_{-0.11}^{+0.18}$\,pc. The closest approach times and distances are different for the two stars despite them likely forming a wide binary due to the large uncertainty on the relative radial positions and velocities, and the fact that this calculation does not incorporate gravitational forces (see Section~\ref{sec:nbody}). We estimate masses of 1.37 and 1.22 $M_{\odot}$ for the two stars based on their spectral type. The closest approach times and distances for the two stars is shown in Figure \ref{fig:2}. These distances are comparable to the closest approach of \object{WISE J072003.20-084651.2} to the Sun 70 \,kya at $0.25^{+0.11}_{-0.07}$ \,pc \citep{Mamajek:2015gg}, although the solar system is significantly more compact than the HD 106906 system. 

HIP 59716 and HIP 59721 are presently separated from each other by $24\arcsec$ on the sky and are listed in the Washington Double Star Catalog \citep{Mason:2001ff} as being a physical binary based on the similarity of their parallax. The {\it Gaia} DR2 astrometry is consistent with this assessment; the parallax of the two stars are indistinguishable ($\Delta \pi=0.025\pm0.053$\,mas), and the proper motion difference ($0.61\pm0.07$\,mas\,yr$^{-1}$) is less than the circular motion of a face-on 2,750\,au orbit ($\sim1.7$\,mas\,yr$^{-1}$). The similar kinematics of this group of stars had been previously noted. Using {\it Hipparcos} astrometry, \citet{Shaya:2010dt} estimated a probability of 92\,\% that HD 106906 and HIP 59716 were bound. No assessment was made regarding the nature of HIP 59721 due to the poor quality of the parallax measurement in the {\it Hipparcos} catalog ($\pi=7.56\pm5.84$\,mas, \citealp{vanLeeuwen:2007dc}).

While the {\it Gaia} astrometry provides exquisite constraints on the position and motion of the two stars in the sky plane (projected separation to within 0.006\,au, relative motion to within 0.008\,au\,yr$^{-1}$), the position and motion in the radial direction is less well constrained; the relative distance is known to within 140,000\,au, and the relative radial velocity to within 0.3\,au\,yr$^{-1}$. This precision is insufficient to confirm that the two stars are bound. Instead, we make a probabilistic argument based on the chance of finding two stars in LCC with a projected separation within 24$\arcsec$ that share a common parallax and proper motion. Using a simplistic assumption that the Galactic positions and motions of LCC members within our sample follow normal distributions, we simulated $10^8$ stars and compared their positions, parallax, and proper motions in the sky plane to that of HIP 59716. Only 20 of the $10^8$ stars were within $24\arcsec$, had a parallax within 0.078\,mas, and had total proper motion within 0.68\,mas\,yr$^{-1}$, implying a chance alignment probability of $\sim$10$^{-7}$.

\subsection{$N$-body simulations}
\label{sec:nbody}
The proximity of HD 106906 and HIP 59716 at closest approach, and between HIP 59716 and HIP 59721 at the present epoch, motivated us to investigate the effect of  the gravitational interaction between the stars and the influence of the Galactic potential. We used the $N$-body \texttt{REBOUND} package \citep{Rein:2012cd,Rein:2015ib} to predict the position of the three stars within a Galactic potential \citep{Bovy:2015gg} over the previous 10\,Myr. We initialized $10^4$ simulations to sample the uncertainties on the astrometry and radial velocity for each star. Each simulation was integrated backwards in time, with the absolute positions and velocities recorded every 100 years. The closest approach times and distances were unchanged from the linear approximation $t_{\rm CA}=-3.5_{-1.8}^{+0.9}$\,Myr, $D_{\rm CA}=0.63_{-0.39}^{+0.92}$\,pc for HIP 59716 and $t_{\rm CA}=-2.2_{-1.0}^{+0.6}$\,Myr, $D_{\rm CA}=0.62_{-0.17}^{+0.21}$\,pc for HIP 59721 (marginalized distributions shown in Figure \ref{fig:2}).

The consistency between the simulations with and without gravity was expected. The difference in the Galactic potential experienced by the three stars over a small fraction of their Galactic orbital period is negligible, and the large uncertainty on the relative radial velocity and distance between HIP 59716 and HIP 59721 meant that the pair were only bound in $\sim$2\,\% of the simulations. The closest approach distances were not significantly different for the bound simulations, increasing slightly for HIP 59716 to $D_{\rm CA}=0.42_{-0.17}^{+0.30}$\,pc, and decreasing slightly for HIP 59721 to $D_{\rm CA}=0.42_{-0.18}^{+0.30}$\,pc. We note that these values are derived from a small number of simulations (212 of 10,000), and may not constitute a representative sample of bound orbit trajectories. The projected separation between the two stars of 2,750\,au suggests a comparable semi-major axis with the $a/\rho$ conversion factor peaking at unity for a uniform eccentricity distribution \citep{Dupuy:2011ip}. Only in the most extreme cases where $a/\rho \sim 3$ and $e\sim 1$ does the apocenter distance, where eccentric binaries spend more of their time, become comparable to the closest approach distance. For the purposes of this study, we assume that the dynamics of the binary system---if it is indeed bound---does not have a large effect on the flyby of HD 106906.

\subsection{Flyby geometry}
The position and motion for the three stars in the Galactic and Galactocentric coordinate systems are given in Table~\ref{tab:1}. We defined two new coordinate systems with HD 106906 at the origin (in terms of both position and velocity). The first has $-X^{\prime}$ pointing toward the current location of the Sun and $Y^{\prime}$ and $Z^{\prime}$ pointing toward East and North from the perspective of the Sun. In this coordinate system, the plane of the disk is rotated $15^{\circ}$ about the $X^{\prime}$ axis and $85^{\circ}$ about the $Y^{\prime}$ axis. The second is rotated  about the first such that the disk lies in one of the planes of the coordinate system ($Y_{\rm disk}Z_{\rm disk}$). A schematic diagram is shown in Figure \ref{fig:3}. 

The trajectories of both stars were almost coplanar with the current plane of the debris disk: $\Delta \theta = 5\fdg4\pm1\fdg7$ ($5\fdg2_{-1.8}^{+1.5}$) for HIP 59716 and $\Delta \theta = 4\fdg2_{-1.1}^{+0.9}$ ($4\fdg8\pm1\fdg0$) for HIP 59721 (Fig. \ref{fig:1}, second column; values in parentheses were calculated from the $N$-body simulation). The relative inclination between the flyby and the orbit of the planet cannot be constrained due to the lack of multi-epoch astrometry to measure the planet’s orbit. We calculated a relative velocity of $3.3\pm1.1$\,km\,s$^{-1}$ and $5.3\pm1.7$\,km\,s$^{-1}$ at the time of closest approach in the $N$-body simulation for HIP 59716 and HIP 59721, respectively. The velocity vector is almost entirely in the disk plane, consistent with a coplanar encounter.

\section{Effect of measurement uncertainties}
\begin{figure}
\epsscale{1.15}
\plotone{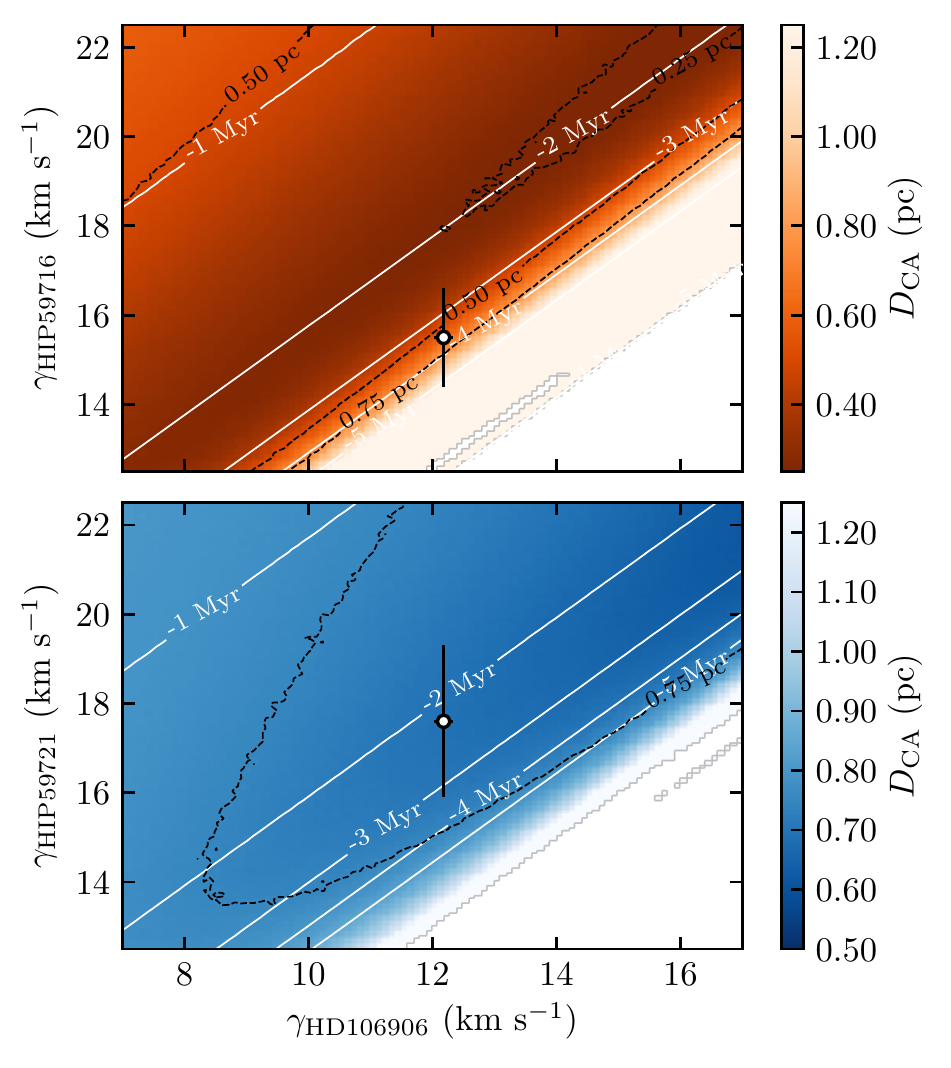}
\caption{Correlation between $D_{\rm CA}$ (color scale, dashed contours), $t_{\rm CA}$ (solid contours), and the relative velocities of HD 106906, HIP 59716 (top panel), and HIP 59721 (bottom panel). The radial velocities used in this study are indicated with error bars denoting 1-$\sigma$ uncertainties. The color scales for each plot have been normalized to the range of $D_{\rm CA}$ values for each star.}
\label{fig:changing_rv}
\end{figure}

The relative three-dimensional motion between HD 106906 and the two candidate perturbers is almost entirely in the radial direction, and as such our uncertainty on the closest approach distance is almost entirely due to the uncertainty in the relative radial positions and motions of the three stars. While the uncertainty in the relative radial positions of the three stars is a factor of $10^8$ larger than the relative tangential positions, the dominant source of uncertainty on the predicted closest approach distance is the relative radial velocity as this error accumulates as the positions of each star are traced back in time. 

To explore the effects of varying the relative radial velocities on the closest approach differences, we repeated the traceback analysis for HIP 59716 and HIP 59721 for a range of plausible radial velocities. For each combination of radial velocities, we drew $10^4$ random variates from the multivariate normal distribution that describe the {\it Gaia} astrometry and their correlation for each star and determined $D_{\rm CA}$ and $t_{\rm CA}$ using the linear approximation described previously. We assume that the radial velocities are perfectly measured. The correlations between $D_{\rm CA}$ and $t_{\rm CA}$ and the radial velocities of the three stars are shown in Figure \ref{fig:changing_rv}. $D_{\rm CA}$ and $t_{\rm CA}$ are only sensitive to the relative radial velocity between HD 106906 and the two other stars rather than their absolute velocities; lines of constant $D_{\rm CA}$ and $t_{\rm CA}$ follow lines of constant relative velocity. Increasing the relative velocity between HD 106906 and HIP 59716 causes a significant decrease in $D_{\rm CA}$ and an increase in $t_{\rm CA}$ (a more recent flyby). Interestingly, if HIP 59716’s radial velocity was similar to that of HIP 59721, the median value of $D_{\rm CA}$ would be near the global minimum found in this grid search. HIP 59716 and HIP 59721 both have relatively large uncertainties on their radial velocities, and periodic measurements will be required in order to rule out the presence of binary companion to either of the two stars.

\section{Magnitude of Dynamical Interaction}
There are two peculiar features of the HD 106906 system that may be explained by the dynamical perturbation caused by a passing star: the asymmetric debris disk \citep{Kalas:2015en,Lagrange:2016bh} and the planet at a large projected separation \citep{Bailey:2014et}. The debris disk has two components: an inner ring at a radius of 50\,au that exhibits a strong brightness asymmetry, and a sharp feature resolved with {\it HST} extending at least 500\,au westward along the disk plane with no counterpart seen on the eastern side \citep{Kalas:2015en}. Several theories have been suggested to explain the observed asymmetry including perturbation by the planet on its current orbit \citep{Jilkova:2015jo,Nesvold:2017ho}, and a scattering event where the planet interacted with either an unseen low-mass companion \citep{Kalas:2015en} or the inner binary \citep{Rodet:2017hr} and then migrated through the disk. The presence of the planet at a projected separation of 738\,au is also a mystery. Canonical planet formation theories suggest that massive planets cannot form at such wide separations. Instead, the planet must have either formed closer in and later migrated or was scattered to its current separation, or formed more like a binary star in the initial fragmentation of the molecular cloud as HD 106906 itself was being formed.

The flyby of HIP 59716 and HIP 59721 may have been important in the context of both the observed asymmetry in the debris disk and the current wide separation of the planet. The two stars passed well within the tidal radius of the HD 106906 binary (1.9\,pc; \citealp{Jiang:2010jk,Mamajek:2013gz}) interior to which the gravitational interaction between the stars is stronger than the Galactic potential. In a recent study, \citet{Rodet:2017hr} postulated that the present orbit of the planet could be explained by a combination of dynamical interaction with an eccentric inner binary (we find $e\sim0.67$ based on our orbit fit discussed previously) and a close encounter with a passing star.

In this scenario, the planet formed at a location within the circumbinary disk that was stable to the dynamical effects of the binary. After formation, the planet migrated inward until it was temporarily trapped in a 1:6 mean-motion resonance with the inner binary. Dynamical interaction with the inner binary cause the semi-major axis of the planet to rapidly oscillate before being ejected from the inner system. \citet{Rodet:2017hr} suggests that the very eccentric (or hyperbolic) orbit of the ejected planet could then be stabilized by an interaction with a passing star that decreases the eccentricity of the orbit and raises the periapsis distance out of the chaotic zone surrounding the binary. Without a stabilizing encounter the planet would enter the chaotic zone surrounding the inner binary each time it goes through periapsis, rapidly leading to the complete ejection of the planet from the system.

The configuration of the flyby required to increase the periapsis distance out of the chaotic zone depends on the mass and closest approach distance of the passing star, the inclination of the encounter, and the angle between the velocity vector of the planet and passing star. Such a scenario has also been suggested for the hypothesized ninth planet within our own solar system \citep{Batygin:2016ef}, with the planet being ejected from the inner solar system due to interactions with the other gas giants, and then stabilized by the gravitational influence of a passing star \citep{Bromley:2016kg}.

In the scenario described by \citet{Rodet:2017hr}, the planet is rapidly ejected from the inner system after reaching the 1:6 mean-motion resonance with the inner eccentric binary. The window for a stabilizing timescale is short ($<$10$^4$ yr); either the planet is ejected immediately on a hyperbolic trajectory, or it first achieves a highly eccentric orbit that evolves into a hyperbolic trajectory after a few passes through the chaotic zone surrounding the binary. Given a mean age of 15\,Myr for LCC members, and an intrinsic scatter of 6\,Myr \citep{Pecaut:2016fu}, the plausible ages for HD 106906 range from 9--21\,Myr. Assuming a younger age of $\sim$10 Myr, the formation of the planet and migration into mean-motion resonance with the inner binary has to occur in $\sim$5\,Myr for the planet to be ejected as the perturbing stars were passing by. These timescales are similar to those proposed for planet formation via pebble accretion (e.g., \citealp{Lambrechts:2014iq}), although formation via gravitational instability would occur much more rapidly (e.g., \citealp{Boss:2011ic}). 

For the older age estimate of $\sim$20 Myr, the planet would need to spend over ten million years outside of the chaotic zone surrounding the binary before entering the mean-motion resonance approximately four million years ago. These timings are inconsistent with the requirement of a massive circumstellar disk to enable planetary migration; disks around the majority of young stars dissipate within a few million years (e.g., \citealp{ansdell17a}). In this case, if a stabilizing encounter did occur, it is more likely to have been with another cluster member in the more distant past, rather than with HIP 59716 and HIP 59721. Future work could search for a dynamical process where the ejection of HD 106906 b occurs at a later epoch than calculated by \citet{Rodet:2017hr}.  

\subsection{Flyby simulations}
\begin{figure}
\epsscale{1.15}
\plotone{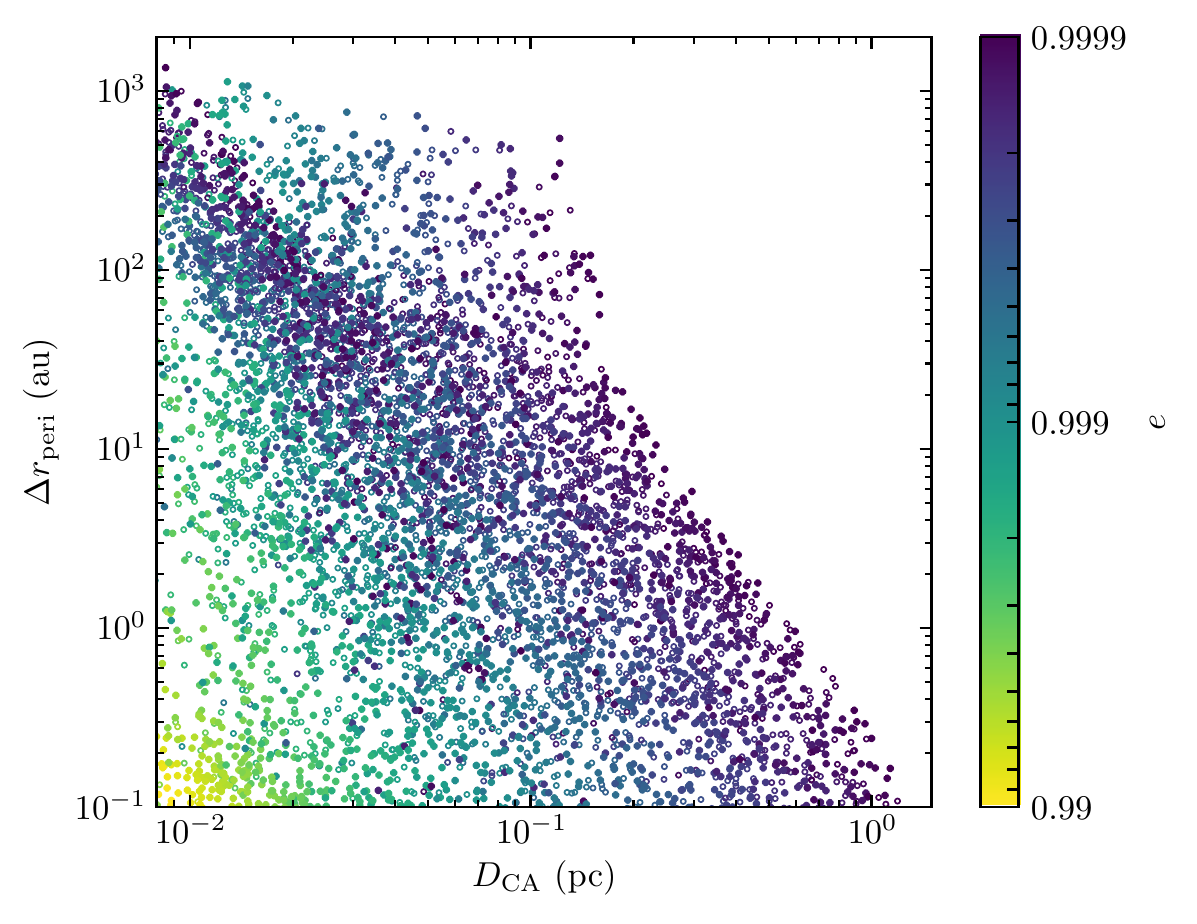}
\caption{The correlation between the change in periapsis distance ($\Delta r_{\rm peri}$) of an orbiting planet as a function of closest approach distance ($D_{\rm CA}$) for the 21,651 prograde (filled symbols) and 6,435 retrograde (open symbols) $N$-body flyby simulations that resulted in a significant increase of $r_{\rm peri}$ for the planet. The symbols are coloured on a logarithmic scale according to the eccentricity of the planet at the start of the simulation.}
\label{fig:flyby_sim}
\end{figure}
We ran $5 \times 10^4$ \texttt{REBOUND} $N$-body simulations to explore the dynamical impact of a stellar flyby on a planet with a highly eccentric ($e>0.9$) orbit. For these simulations we assumed that HIP 59716 and HIP 59721 form a physical binary, and treated them as a single particle of their combined mass given current uncertainties about their orbit. This is a reasonable approximation when the flyby distance is significantly larger than the binary separation, where the gravitational influence of the two stars would not be resolved by the orbiting planet, but may not be valid in cases where the minimum separation between the planet and perturbing stars is comparable to their orbital semi-major axis. We note that these simulations do not model the ejection mechanism itself, rather the planet is initialized on a highly eccentric orbit to simulate the orbital configuration after an ejection has occurred. A more detailed modeling effort will be needed to combine an ejection event with a stellar flyby into a single simulation.

Each simulation was initialized with a 2.71\,$M_{\odot}$ particle at $(0, 0)$\,pc, accounting for the combined mass of the HD 106906 binary, around which a 11\,$M_{\rm Jup}$ particle was placed with an orbital eccentricity of $\log_{10}(1-e)$ drawn from $\mathcal{U}(-4, -1)$ (where $\mathcal{U}$ denotes the uniform distribution), a semi-major axis with a corresponding periapsis distance ($r_{\rm peri}$) of 1.5\,au, and a mean anomaly $M$ drawn from $ \mathcal{U}(0, 2\pi)$\,rad so that the flyby occurs at a different phase of the orbit within each simulation. The two candidate perturbers were modeled by a single particle with a mass of 2.59\,$M_{\odot}$, initialized at $y = -3$\,pc, $\log_{10}(x)$ drawn from $\mathcal{U}(-2, 0.5)$\,pc, with a velocity of $dy/dt = 3$\,km\,s$^{-1}$. In half of the simulations the perturber was instead initialized at $y = 3$\,pc, with a velocity of $dy/dt = -3$\,km\,s$^{-1}$, to account for the unknown direction of the orbit of the planet. A coplanar encounter was assumed to limit the size of phase space to explore. Each simulation was advanced either until the perturbers had reached $y = 3$\,pc (or $-3$\,pc), or for 2\,Myr, whichever occurred first.

Of the 25,000 prograde (retrograde) flybys; 21,670 (6,385) caused an increase in $r_{\rm peri}$ and 3,330 (18,615) led to a decrease in $r_{\rm peri}$, with 1,541 (1,525) resulting in the planet being ejected from the system. The correlation between the change in periapsis distance ($\Delta r_{\rm peri}$) and the closest approach distance between the two stars for the simulations in which the eccentricity decreased is shown in Figure \ref{fig:flyby_sim}. There is a clear correlation between $\Delta r_{\rm peri}$ and $D_{\rm CA}$, with closer encounters leading to a larger $\Delta r_{\rm peri}$ for a given initial eccentricity. The presence of a debris ring at $\sim$50\,au around the HD 106906 binary suggests that the $r_{\rm peri}$ for the planet is now $>50$\,au, otherwise the disk would be disrupted with each periastron passage for relative inclinations $\lesssim10$\,deg \citep{Jilkova:2015jo}. Periapsis distances interior to the radius of the debris ring are possible for higher relative inclinations ($\gtrsim40$\,deg) without disrupting the disk. These $N$-body simulations suggest that $r_{\rm peri}$ for a highly eccentric planet can be plausibly raised from $\sim$1.5 to $\sim$100\,au with a co-planar stellar flyby within 0.2\,pc. Future observational and theoretical work is needed to more precisely constrain the geometry of the encounter and its dynamical consequences.

\section{Conclusions}
HIP 59716 and HIP 59721 are the best candidates of the currently known members of Sco--Cen for a dynamically important close encounter with HD 106906 within the last 15\,Myr. The flyby of these two stars fulfill many of the criteria for the stabilization scenario described in \citet{Rodet:2017hr}. Their trajectories are almost coplanar with the debris disk in its current orientation, their velocities relative to HD 106906 at closest approach are low (the change in velocity of the orbiting planet being inversely proportional to the relative velocity of the passing star at closest approach), and the distribution of closest approach distances for HIP 59716 is consistent with a dynamically significant encounter within 0.5\,pc.

The biggest source of uncertainty in the closest approach distances are the relative radial velocities. Future spectroscopic observations of the two candidate perturbers will be essential to precisely determine their radial velocities. For example, increasing the relative velocity between HD 106906 and HIP 59716 by 1\,km\,s$^{-1}$ significantly increases the probability of a closest approach distance within 0.1\,pc (Fig. \ref{fig:changing_rv}). The astrometry for each star will also be improved with  upcoming {\it Gaia} data releases. Not only will the precision of the measurements improve, but the photocenter motion of short period binaries will also be accounted for. With these data in hand, a more precise determination of the kinematics of the three stars can be made.

\acknowledgments
We wish to thank Eric Nielsen, Ian Czekala, Ruth Murray-Clay, and Anne-Marie Lagrange  for useful discussions relating to this work. We also wish to thank the referee for the comments that helped improve the quality of this work. The authors were supported in part by NSF AST-1518332, NASA NNX15AC89G and NNX15AD95G. This work benefited from NASA's Nexus for Exoplanet System Science (NExSS) research coordination network sponsored by NASA’s Science Mission Directorate. This work has made use of data from the European Space Agency (ESA) mission {\it Gaia} (\url{https://www.cosmos.esa.int/gaia}), processed by the {\it Gaia} Data Processing and Analysis Consortium (DPAC; \url{https://www.cosmos.esa.int/web/gaia/dpac/consortium}). Funding for the DPAC has been provided by national institutions, in particular the institutions participating in the {\it Gaia} Multilateral Agreement. This research has made use of the SIMBAD database and the VizieR catalog access tool, both operated at the CDS, Strasbourg, France.

\software{Astropy \citep{TheAstropyCollaboration:2013cd},  
          Matplotlib \citep{Hunter:2007ih},
          galpy \citep{Bovy:2015gg},
          REBOUND \citep{Rein:2012cd}}

\bibliography{106906}
\end{document}